# A Comprehensive Survey on Bengali Phoneme Recognition

**Sadia Tasnim Swarna** [1*], **Shamim Ehsan** [1], **Md. Saiful Islam**[1] **and Marium E Jannat** [1]

[1] Department of Computer Science and Engineering, Shahjalal University of Science & Technology, Sylhet, Bangladesh.

Email: sadiatasnimswarna@gmail.com, ehsanhrid@gmail.com, saiful-cse@sust.edu, jannatcse@yahoo.com.

**Abstract**: Hidden Markov model based various phoneme recognition methods for Bengali language is reviewed. Automatic phoneme recognition for Bengali language using multilayer neural network is reviewed. Usefulness of multilayer neural network over single layer neural network is discussed. Bangla phonetic feature table construction and enhancement for Bengali speech recognition is also discussed. Comparison among these methods is discussed.

## 1. INTRODUCTION

A phoneme is one of the units of sound (or motion on account of gesture-based communications) that differentiate single word from another in a specific dialect. At the point when two sounds can be utilized to separate words, they can be said to have a place with various phonemes. Barman (2011) says that a phoneme is portrayed by certain particular components which make it a different element in an arrangement of sounds available in a language. Although Bengali is one of the largely spoken languages on the planet, yet not very many literature have been found in automatic speech recognition (ASR) for the Bengali language. More than 265 million individuals talk in Bengali as their local dialect. A major difficulty to research in Bangla ASR is the lack of proper speech corpus though some efforts are made to develop Bangla speech corpus to build a Bangla text to speech system is stated in "Experiments with unit selection speech databases for Indian languages." By Kishore, S. P., et al.( 2003) The aim of this paper is to give a survey of automatic phoneme recognition for Bengali language using distinctive strategies. More emphasis will be put on phonetic feature enhancement for Bangla ASR as it gives higher sentence correct rate (SCR), word correct rate (WCR) and word accuracy (WA) contrasted with alternate techniques.

### 1.1 Bangla phoneme

There are 14 vowels and 29 consonants in the phonetic inventory of Bengali language. Vowel set includes seven nasalized vowels. In IPA an approximate phonetic scheme is given in Ghulam at el.(2009) in his "*Automatic speech recognition for Bangla digits"*, and C. Masica(1991) in his book "Indo-Aryan Languages". Here only the main 7 vowel sounds are shown, though there is two more long counterparts of /i/ and /u/ denoted as /i:/ and /u:/ respectively.

---

* Corresponding author: sadiatasnimswarna@gmail.com

*1.2 Scope of Automatic Speech Recognition or Phoneme Recognition:*

The majority of ASR systems today do incorporate a phonetic level of representation. There's a variety of reasons, probably the most relevant one is that it represents a useful level of abstraction that allows training data to be used effectively (that is, there are probably too many states below that level and it's difficult to get representative samples, e.g. of words, above that level). In decoding search, it provides a useful source of constraint.

On a more practical level, the point of ASR is to transcribe language (i.e. words), so the focus is on how to combine the different levels to produce accurate word results. Phonetic transcription, per se, is not particularly a focus.

Phonetic transcription does have a role in systems that try to deal with out-of-vocabulary (OOV) words, which by definition do not appear in the vocabulary. Phonetic (and word-fragment) decoding provides a mechanism by which an ASR system can "learn" new words and incorporate them into subsequent decoding.

## 2. PHONEME RECOGNITION METHODS

The developments of various kinds of hidden Markov model (HMM) based automatic speech recognition systems can be found in recent works done by .Hossain, el al. (2007) on their "Bangla vowel characterization based on analysis by synthesis.", Hasnat et al. (2007) on their Isolated and continuous Bangla speech recognition: implementation, performance and application perspective." paper, Hoque and A. K. M(2006) on their "Bengali Segmented automated speech recognition.", Kirchhoff al el. (2002) on their "Combining acoustic and articulatory feature information for robust speech recognition. And others where Bengali automatic speech recognition are found some extent. A preprocessed form is used most of the time by these ASR systems, like Mel-frequency cepstral coefficients (MFCCs) of the speech signal, which encodes the time-frequency distribution of signal energy. However, in real acoustic conditions, these MFCC based systems do not provide better recognition performance. On the contrary, an articulatory feature based system (AFs) or phonetic features (PFs) exhibits a higher recognition which is accurate in practical conditions, and models co-articulatory phenomenon more naturally. Nahid, Md Mahadi Hasan et al.(2016) on their "A Noble Approach for Recognizing Bangla Real Number Automatically Using CMU Sphinx4" proposed a method for automatic Bangla Real Number recognizer using the speech recognizer API CMU Sphinx toolkit. They have also used MFCC feature extraction methods for phoneme recognition which is needed for the intermediate process for recognizing Bangla Real Numbers.

*2.1 Multilayer neural network:*

Bengali phoneme recognition utilizing Artificial Neural Network (ANN) is reported in the paper of Roy et el. (2002) named "Development of the speech recognition system using artificial neural network.". Be that as it may most of these works use a very small database and are focused on simple recognition or on the frequency distributions of different vowels and consonants. Another work related to multilayer neural network is done by Kotwal and Alam (2010). Their technique utilizes multilayer neural network (MLN) as single layer neural network can't resolve coarticulation impact. An exact phoneme recognizer usefulness is required as present Hidden Markov Model (HMM) based automatic speech recognition systems give some error for new out-of-vocabulary (OOV) word. This method by Kotwal and Alam utilizes standard back-propagation algorithm for training multilayer neural network. Single MLN has a failure to model dynamic information precisely.

This technique has two stages, Firstly, a multilayer neural network converts acoustic features named Mel-frequency cepstral coefficients (MFCCs) into phoneme probabilities of 39 dimensions. Secondly, $\Delta$ and $\Delta\Delta$ parameters for corresponding phoneme probabilities calculated by linear regression (LR) and it is inserted into hidden Markov model (HMM) based classifier.

Combining output probabilities and $\Delta$ and $\Delta\Delta$, input features are found. 39 dimensions are (12-MFCC, 12-$\Delta$MFCC, 12- $\Delta\Delta$MFCC, P, $\Delta$P and $\Delta\Delta$, here P refers raw log energy of the input speech signal).

Three experiments are designed for obtaining PCR and PA
        i)      MFCC+MLN [9]
        ii)     MFCC+MLN+ Δ
        iii)    MFCC+MLN+ Δ. ΔΔ (proposed)

Figure 1 consists the flow chart of their proposed phoneme recognition method:

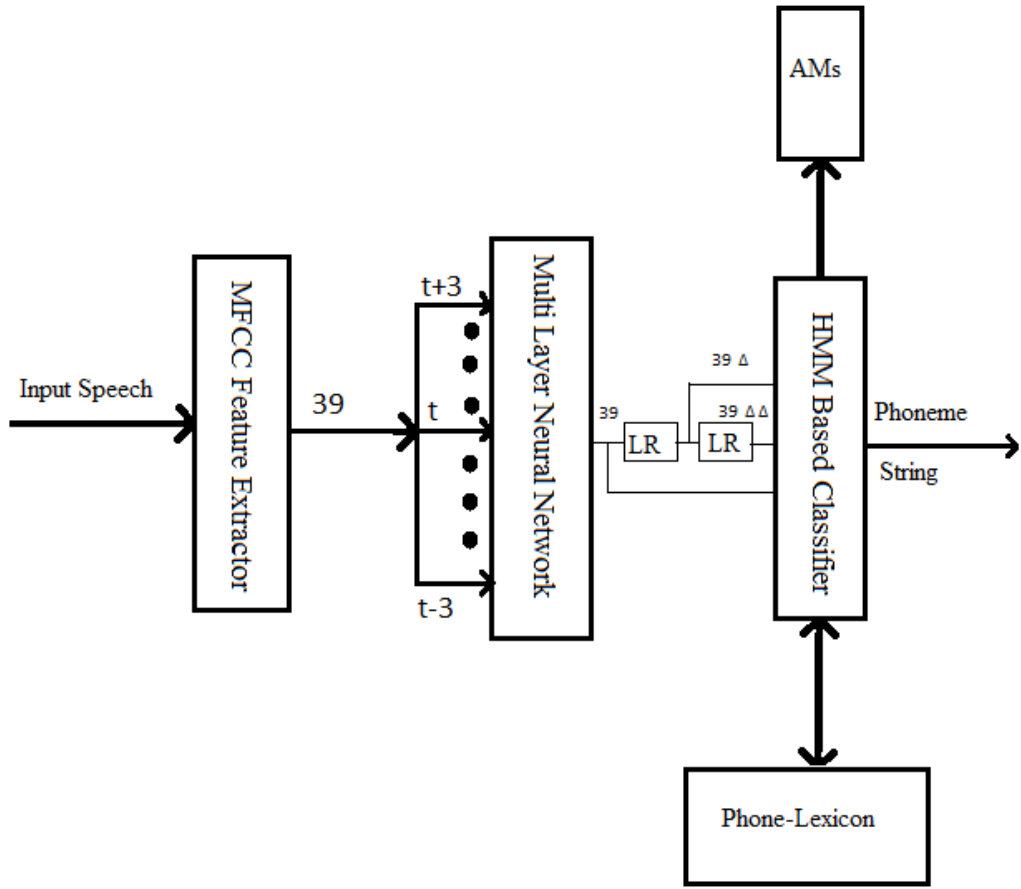

Figure 1: Phoneme recognition method using Multilayer ANN.

## 2.2 Phonetic Feature Table:

With its phonetic features, a phoneme can be identified specifically, as stated by S. King in his "*Speech recognition via phonetically features syllables*" (1998) paper. There are in total 53 phonemes in Bangla language. The phones, (/s/) and (/s/), and, (/n/) and (/n/) share same Phonetic features and contain approximately same spectrum. Twenty-two phonetic features are needed for representing all the Bangla phonemes. Twenty-two PFs for each phoneme are silence, a short silence, stop, nasal, bilabial, fricative, liquid, lenis, vowel, front, central, back, unvoiced, long, short, diphthong, high, low, medium, round, unround and glottal.

Hassan el al. (2013) in their paper "Bangla phonetic feature table construction for automatic speech recognition" have implemented an ASR system based on phonetic features with an input acoustic vector of MFCCs using an

MLN. This system consists of three steps; firstly, acoustic features, mel frequency cepstral coefficients (MFCCs) extraction is done. Then it embeds PFs extraction procedure using a multilayer neural network (MLN), Finally, in the last stage integrates a triphone-based HMM for generating the output text strings by inputting logarithmic values of twenty-two dimensional PFs.

By comparing the performance of standard MFCC-based Method with their proposed method, they have designed two experiments:
i) MFCC:dim-39 [Baseline]
ii) PF:dim-22 [Proposed]

39-dimensional MFCCs and log values of 22-dimensional PFs are input features for the classifier. Experiments are done on these mixture components 1, 2, 4 and 8 respectively.

### *2.3 Phonetic Features Enhancement:*

A research by Kabir and Rasel (2015) proposes a phonetic feature (PF) extraction method for Bangla continuous word recognition with high performance in an automatic speech recognition (ASR) system. The method consists of three stages: first one is a multilayer neural network (MLN) is used for mapping continuous acoustic features, local features (LFs) onto PFs. Secondly, it incorporates inhibition/enhancement (In/En) functionalities to discriminate whether the PF dynamic patterns of trajectories are convex or concave. Patterns are enhanced for convex patterns and inhibited for concave patterns. Then it normalizes the PF vector using Gram-Schmidt algorithm and then passes through a Hidden Markov Model (HMM) based classifier. This feature extraction method provides higher sentence correct rate (SCR), word correct rate (WCR) and word accuracy (WA) compared to the methods that not incorporated In/En network They designed the following experiments for evaluating the performance of the standard MFCC-based method. Their incorporated method was also included.

a) MFCC:dim-39 [Baseline]
b) PF(MFCC) + log10:dim-22
c) PF(MFCC)+ Norm:dim-22
d) PF(LF)+Norm:dim-22
e) PF(MFCC)+In/En+Norm:dim22
f) PF(LF)+In/En+Norm:dim-22 (Proposed)

### 3. RESULT AND ANALYSIS

In their approach of Kotwal and Alams' multi-layer ANN, they used dynamic parameters and get a better result than the current strategies. Their proposed strategy gives higher phoneme and sentence correct rate than the previous methods investigated. They have 53.5%, 56.50%, 59.70% and 70.78% percent phoneme correct rate while the existing method in that period has 53%, 57.50%, 58.25% 70.37% phoneme correct rate for the mixture components of 1,2,4 and 32 respectively.

On account of ASR utilizing phonetic feature table, their proposed method (PF:dim-22) have produced a better result than the baseline system MFCC:dim-39. The results of word correct rate and sentence correct rate are shown in figure 2.

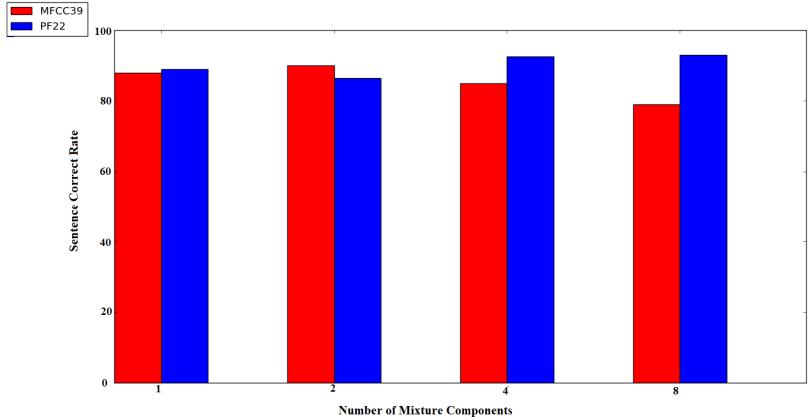

Fig 2: Sentence Correct Rate of MFCC32 and the proposed method of PF table

From figure, we can see for mixture components of one, two, four and eight, correct rate of sentences is 89.20%, 88.20%, 91.50% and 91.60% while baseline system with MFCC32 model generates 88.60%, 90.00%, 85.00% and 79.20% for similar test.

For the instance of ASR utilizing phonetic feature enhancement for mixture components one, two, four and eight, correct rate of sentences is 97.70%, 97.90%, 97.00% and 97.90% where with MFCC39 demonstrate produces 88.60%, 88.20%, 91.50% and 91.60% for same test dataset used. A theoretical comparison between Multilayer ANN, Phonetic Feature Table and Phonetic Feature Enhancement are shown on both figure 4 below.

A table is also included to get a better understanding about the comparisons.

Table 1: Performance comparison of different phoneme recognition methods:

| Mixture Components | Multilayer ANN | Phonetic Feature Table | Phonetic Feature Enhancement |
|---|---|---|---|
| 1 | 53.5 | 89.20 | 97.70 |
| 2 | 56.5 | 88.20 | 97.90 |
| 4 | 59.70 | 91.50 | 97.00 |
| 8 | - | 91.60 | 97.90 |
| 32 | 70.76 | - | - |

From the table and figure we can see that phonetic feature enhancement technique is superior to anything the past two strategies examined in this paper when the same data set is used.

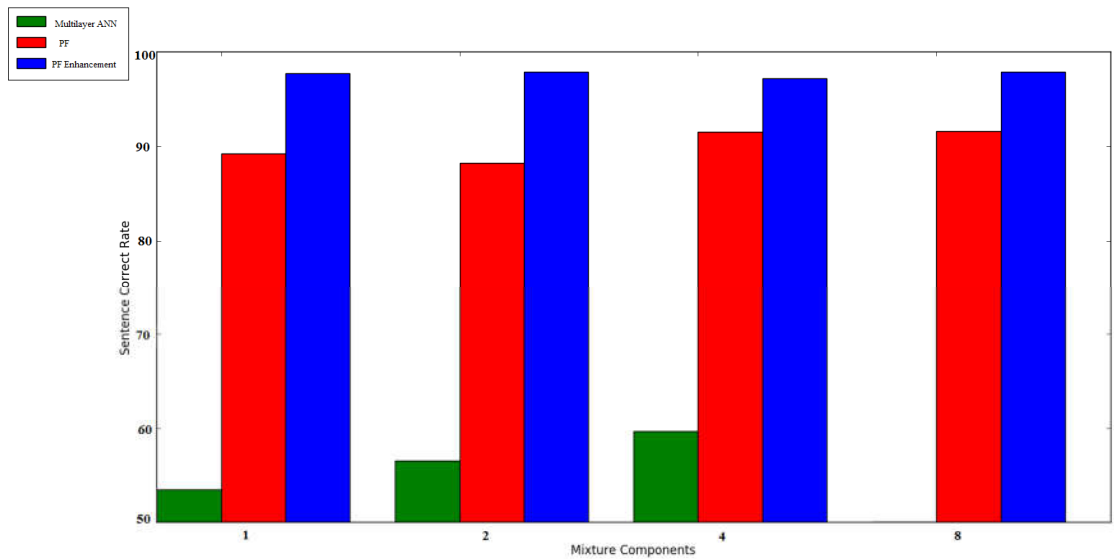

Figure 3: Sentence Correct Rate doe different Phoneme Recognition Methods.

## 4. CONCLUSION

The aim of this paper was to give a review of different approaches of phoneme recognitions to its readers. Multilayer neural network based phoneme recognition, phonetic feature table construction based ASR and phonetic feature enhancement based were discussed. The different advantage of three phoneme recognition methods has been pointed out.